\relax
\relax

\relax

\documentclass[a4paper,onecolumn,superscriptaddress,11pt]{quantumarticle}

\relax

\pdfoutput=1
\usepackage{amsfonts,amsmath}
\usepackage{color}
\usepackage{graphicx}
\usepackage[arrow,curve,matrix,frame]{xy}
\usepackage[utf8]{inputenc}
\usepackage[T1]{fontenc}

\definecolor{darkgreen}{rgb}{0,0.4,0}

\def\mypdfdest#1{\leavevmode\vbox to 0pt{\vss \pdfdest name{#1}XYZ\relax\vskip\baselineskip}}
\catcode`@=11
\catcode`\^^A=3
\def\showurl#1{{\edef\theuri{#1}\pdfstartlink attr{/Border[0 0 0]/H/I/C[0.3 0.3 1]}user{/Subtype/Link/A<</Type/Action/S/URI/URI(\theuri)>>}\relax
\textcolor{blue}{\tt\hyphenchar\font=-1\def\nextpenalty{}\def\next{}\dosplit#1^^A^^A}\pdfendlink}}
\def\splittt#1{{\tt\hyphenchar\font=-1\def\nextpenalty{}\def\next{}\dosplit#1^^A^^A}}
\def\dosplit{\ifx\next^^A\else
\ifx\next/{/\gdef\nextpenalty{\penalty0}}\else
\ifx\next.{.\gdef\nextpenalty{\penalty50}}\else
\ifx\next:{:\gdef\nextpenalty{\penalty50}}\else
\ifx\next-{-\gdef\nextpenalty{\penalty50}}\else
\ifx\next?{?\gdef\nextpenalty{\penalty50}}\else
\ifx\next={=\gdef\nextpenalty{\penalty50}}\else
\ifx\next##{\char'043\gdef\nextpenalty{\penalty50}}\else
\ifx\next&{\char'046\gdef\nextpenalty{\penalty50}}\else
\ifx\next_{\char'137\gdef\nextpenalty{\penalty50}}\else
\ifx\next({\penalty50(\gdef\nextpenalty{}}\else
\ifx\next~{\penalty50\char'176\gdef\nextpenalty{}}\else
{\nextpenalty\next\gdef\nextpenalty{}}\fi\fi\fi\fi\fi\fi\fi\fi\fi\fi\fi\afterassignment\dosplit
\fi\let\next=}
\def\doilink#1{{\edef\theuri{https://dx.doi.org/#1}\pdfstartlink attr{/Border[0 0 0]/H/I/C[0.3 0.3 1]}user{/Subtype/Link/A<</Type/Action/S/URI/URI(\theuri)>>}\relax
\textcolor{blue}{#1}\pdfendlink}}

\DeclareRobustCommand\mycite{\@ifnextchar [{\@tempswatrue\mycitex}{\@tempswafalse\mycitex[]}}
\let\cite\mycite
\def\@mycitex[#1]#2{\leavevmode
\let\@mycitea\@empty
\@mycite{\@for\@myciteb:=#2\do
{\@mycitea\def\@mycitea{,\penalty\@m\ }\edef\@myciteb{\expandafter\@firstofone\@myciteb\@empty}\if@filesw\immediate\write\@auxout{\string\citation{\@myciteb}}\fi
\@ifundefined{b@\@myciteb}{\hbox{\reset@font\bfseries ?}\G@refundefinedtrue
\@latex@warning
{Citation
`\@myciteb'
on
page
\thepage
\space
undefined}}{\@mycite@ofmt{\csname
b@\@myciteb\endcsname}}}}{#1}}
\def\@mycite#1#2{[#1\if@tempswa , #2\fi]}
\let\@mycite@ofmt\hbox

\newcount\backciteanchor
\def\mycitex[#1]#2{\relax
\message{mycitex}\relax
\mypdfdest{citing-\the\backciteanchor}\relax
\protect\@mycitex[#1]{#2}\relax
\immediate\write\@auxout{\string\mybackcite{#2}{\string\S\relax\pdfstartlink
attr{/Border[0
0
0]/H/I/C[1
0
0]}
goto
name{citing-\the\backciteanchor}\string\textcolor{red}{\@currentlabel}\pdfendlink}}\relax
\global\advance\backciteanchor1\relax
}
\global\let\@citex\mycitex
\def\bibcite#1#2{\@newl@bel{b}{#1}{\pdfstartlink attr{/Border[0 0 0]/H/I/C[0 0.4 0]}goto
name{bib-#1}\textcolor{darkgreen}{\bf
#2}\pdfendlink}}
\def\mybackcite#1#2{\relax
\expandafter\ifx\csname mybackcite@#1\endcsname\relax
\expandafter\protected@xdef\csname mybackcite@#1\endcsname{.
Citations
in
this
document:
#2}\relax
\else\relax
\expandafter\protected@xdef\csname mybackcite@#1\endcsname{\csname mybackcite@#1\endcsname, #2}\relax
\fi\relax
}
\def\mybackbibcite#1#2{\relax
\expandafter\ifx\csname mybackbibcite@#1\endcsname\relax
\expandafter\protected@xdef\csname mybackbibcite@#1\endcsname{.
See
#2}\relax
\else\relax
\expandafter\protected@xdef\csname mybackbibcite@#1\endcsname{\csname mybackbibcite@#1\endcsname, #2}\relax
\fi\relax
}
\catcode`@=12

\numberwithin{table}{section}
\numberwithin{figure}{section}

\makeatletter
\renewcommand\labelitemi{$\m@th\bullet$}
\let\c@figure\c@subsection
\let\c@table\c@subsection
\makeatother

\begin{document}

\pdfpagewidth=210 true mm
\pdfpageheight=297 true mm

\title
[Is
the
security
of
quantum
cryptography
guaranteed
by
the
laws
of
physics?]
{Is
the
security
of
quantum
cryptography\\guaranteed
by
the
laws
of
physics?}

\date{2018.03.12}

\relax
\author{Daniel J. Bernstein}
\affiliation{Department of Computer Science, University of Illinois at Chicago, Chicago, IL 60607--7045, USA}
\email{djb@cr.yp.to}
\homepage{https://cr.yp.to/djb.html}
\relax

\relax
\thanks{This work was supported
by
the
European
Commission
under
Contract
ICT-645622
PQCRYPTO;
by
the
Netherlands
Organisation
for
Scientific
Research
(NWO)
under
grant
639.073.005;
and
by
the
U.S.
National
Science
Foundation
under
grant
1314919.
``Any
opinions,
findings,
and
conclusions
or
recommendations
expressed
in
this
material
are
those
of
the
author(s)
and
do
not
necessarily
reflect
the
views
of
the
National
Science
Foundation''
(or,
obviously,
other
funding
agencies).
Permanent
ID
of
this
document:
{\tt
e39346ac8e0f20edb8df3334f8751ac75a600099}.}
\relax

\maketitle

\makeatletter
\def\@currentlabel{1}
\makeatother

\begingroup

\parindent 0pt

\bigskip
\bigskip
\hrule
\medskip
{\it
QKD~\dots~offers
the
ultimate
security
of
the
inviolability
of
a
law
of
Nature
for
key
distribution.}
\hfill ---Hughes,~Alde,~Dyer,~Luther,~Morgan,~and~Schauer,~1995~\cite{1995/hughes}\relax
\medskip
\hrule
\medskip
{\it
Quantum
cryptography
differs
from
conventional
cryptography
in
that
the
data
are
kept
secret
by
the
properties
of
quantum
mechanics,
rather
than
the
conjectured
difficulty
of
computing
certain
functions.}
\hfill ---Shor~and~Preskill,~2000~\cite{2000/shor}\relax
\medskip
\hrule
\medskip
{\it
Quantum
key
distribution
provides
perfect
security
because,
unlike
its
classical
counterpart,
it
relies
on
the
laws
of
physics.}
\hfill ---Christandl, Renner, and Ekert, 2004 \cite{2004/christandl}\relax
\medskip
\hrule
\medskip
{\it
However,
in
contrast
to
these
public
key
schemes,
QKD
as
a
cryptographic
primitive
offers
security
that
is
guaranteed
by
the
laws
of
physics.}
\hfill ---Campagna,
Chen,
Dagdelen,
Ding,
\endgraf\hfill
Fernick,
Gisin,
Hayford,
Jennewein,
Lütkenhaus,
Mosca,
Neill,
Pecen,
\endgraf\hfill
Perlner,
Ribordy,
Schanck,
Stebila,
Walenta,
Whyte,
and
Zhang,
2015
\cite{2015/campagna}\relax
\medskip
\hrule
\medskip
{\it
Quantum
cryptography
solves
the
problem
of
key
distribution
by
allowing
the
exchange
of
a
cryptographic
key
between
two
remote
parties
with
absolute
security,
guaranteed
by
the
fundamental
laws
of
physics.}
\hfill ---ID Quantique, 2016 \cite{2016/idq-understanding}\relax
\medskip
\hrule
\bigskip
\bigskip

\endgroup

\begin{abstract}
It
is
often
claimed
that
the
security
of
quantum
key
distribution
(QKD)
is
guaranteed
by
the
laws
of
physics.
However,
this
claim
is
content-free
if
the
underlying
theoretical
definition
of
QKD
is
not
actually
compatible
with
the
laws
of
physics.
This
paper
observes
that
(1)
the
laws
of
physics
pose
serious
obstacles
to
the
security
of
QKD
and
(2)
the
same
laws
are
ignored
in
all
QKD
``security
proofs''.

\textbf{Keywords:}
security
failures,
quantum
cryptography,
quantum
key
distribution,
QKD,
side-channel
attacks,
electromagnetism,
gravity,
information
flow,
holographic
principle.
\end{abstract}

\definecolor{darkgreen}{rgb}{0,0.4,0}

\def\mypdfdest#1{\leavevmode\vbox to 0pt{\vss \pdfdest name{#1}XYZ\relax\vskip\baselineskip}}
\catcode`@=11
\catcode`\^^A=3
\def\showurl#1{{\edef\theuri{#1}\pdfstartlink attr{/Border[0 0 0]/H/I/C[0.3 0.3 1]}user{/Subtype/Link/A<</Type/Action/S/URI/URI(\theuri)>>}\relax
\textcolor{blue}{\tt\hyphenchar\font=-1\def\nextpenalty{}\def\next{}\dosplit#1^^A^^A}\pdfendlink}}
\def\splittt#1{{\tt\hyphenchar\font=-1\def\nextpenalty{}\def\next{}\dosplit#1^^A^^A}}
\def\dosplit{\ifx\next^^A\else
\ifx\next/{/\gdef\nextpenalty{\penalty0}}\else
\ifx\next.{.\gdef\nextpenalty{\penalty50}}\else
\ifx\next:{:\gdef\nextpenalty{\penalty50}}\else
\ifx\next-{-\gdef\nextpenalty{\penalty50}}\else
\ifx\next?{?\gdef\nextpenalty{\penalty50}}\else
\ifx\next={=\gdef\nextpenalty{\penalty50}}\else
\ifx\next##{\char'043\gdef\nextpenalty{\penalty50}}\else
\ifx\next&{\char'046\gdef\nextpenalty{\penalty50}}\else
\ifx\next_{\char'137\gdef\nextpenalty{\penalty50}}\else
\ifx\next({\penalty50(\gdef\nextpenalty{}}\else
\ifx\next~{\penalty50\char'176\gdef\nextpenalty{}}\else
{\nextpenalty\next\gdef\nextpenalty{}}\fi\fi\fi\fi\fi\fi\fi\fi\fi\fi\fi\afterassignment\dosplit
\fi\let\next=}
\def\doilink#1{{\edef\theuri{https://dx.doi.org/#1}\pdfstartlink attr{/Border[0 0 0]/H/I/C[0.3 0.3 1]}user{/Subtype/Link/A<</Type/Action/S/URI/URI(\theuri)>>}\relax
\textcolor{blue}{#1}\pdfendlink}}

\DeclareRobustCommand\mycite{\@ifnextchar [{\@tempswatrue\mycitex}{\@tempswafalse\mycitex[]}}
\let\cite\mycite
\def\@mycitex[#1]#2{\leavevmode
\let\@mycitea\@empty
\@mycite{\@for\@myciteb:=#2\do
{\@mycitea\def\@mycitea{,\penalty\@m\ }\edef\@myciteb{\expandafter\@firstofone\@myciteb\@empty}\if@filesw\immediate\write\@auxout{\string\citation{\@myciteb}}\fi
\@ifundefined{b@\@myciteb}{\hbox{\reset@font\bfseries ?}\G@refundefinedtrue
\@latex@warning
{Citation
`\@myciteb'
on
page
\thepage
\space
undefined}}{\@mycite@ofmt{\csname
b@\@myciteb\endcsname}}}}{#1}}
\def\@mycite#1#2{[#1\if@tempswa , #2\fi]}
\let\@mycite@ofmt\hbox

\newcount\backciteanchor
\def\mycitex[#1]#2{\relax
\message{mycitex}\relax
\mypdfdest{citing-\the\backciteanchor}\relax
\protect\@mycitex[#1]{#2}\relax
\immediate\write\@auxout{\string\mybackcite{#2}{\string\S\relax\pdfstartlink
attr{/Border[0
0
0]/H/I/C[1
0
0]}
goto
name{citing-\the\backciteanchor}\string\textcolor{red}{\@currentlabel}\pdfendlink}}\relax
\global\advance\backciteanchor1\relax
}
\global\let\@citex\mycitex
\def\bibcite#1#2{\@newl@bel{b}{#1}{\pdfstartlink attr{/Border[0 0 0]/H/I/C[0 0.4 0]}goto
name{bib-#1}\textcolor{darkgreen}{\bf
#2}\pdfendlink}}
\def\mybackcite#1#2{\relax
\expandafter\ifx\csname mybackcite@#1\endcsname\relax
\expandafter\protected@xdef\csname mybackcite@#1\endcsname{.
Citations
in
this
document:
#2}\relax
\else\relax
\expandafter\protected@xdef\csname mybackcite@#1\endcsname{\csname mybackcite@#1\endcsname, #2}\relax
\fi\relax
}
\def\mybackbibcite#1#2{\relax
\expandafter\ifx\csname mybackbibcite@#1\endcsname\relax
\expandafter\protected@xdef\csname mybackbibcite@#1\endcsname{.
See
#2}\relax
\else\relax
\expandafter\protected@xdef\csname mybackbibcite@#1\endcsname{\csname mybackbibcite@#1\endcsname, #2}\relax
\fi\relax
}
\catcode`@=12

\section{The ``provable security'' of quantum cryptography}

The
core
advertisement
for
quantum
cryptography---in
particular,
for
quantum
key
distribution,
which
I'll
focus
on---is
the
claim
that
its
security
is
guaranteed
by
the
laws
of
physics.
One
would
expect
this
claim
to
be
backed
by
a
clearly
stated
theorem
having
the
following
shape:
\begin{itemize}
\item ``Assume $L$.'' Here $L$ is a statement of the laws of physics.
\item ``Assume $P$.'' Here $P$ states physical actions carried out by Alice and Bob.
\item ``Then $S$.'' Here $S$ states a security property:
e.g.,
something
about
the
randomness,
from
Eve's
perspective,
of
a
key
shared
between
Alice
and
Bob.
\end{itemize}
This
theorem
statement
would
provide
a
starting
point
for
security
auditors
to
dive
into
questions
of
whether
the
stated
$L$
is
actually
how
the
real
world
works;
whether
the
stated
$P$
matches
what
is
actually
being
sold
as
``quantum
key
distribution'';
whether
the
proof
of
the
theorem
is
correct;
and
whether
the
stated
$S$
actually
includes
what
the
users
want.

An
auditor
who
attempts
to
find
this
security
theorem
in
the
literature
will
easily
find
papers
claiming
to
present
``security
proofs''
for
several
different
types
of
quantum
cryptography;
see,
e.g.,
\cite{2004/gottesman}.
However,
the
theorems
in
these
papers
never
seem
to
explicitly
hypothesize
$L$,
the
laws
of
physics.

Is
this
merely
a
culture
gap---when
physicists
say
``Theorem''
they
implicitly
mean
a
theorem
under
certain
well-known
hypotheses?
Or
is
there
an
important
reason
that
these
papers
aren't
hypothesizing
Newton's
law
of
gravitation,
for
example,
and
Maxwell's
equations
for
electromagnetism?
Or
refinements
such
as
general
relativity
and
quantum
electrodynamics?

An
auditor
who
dives
more
deeply
into
the
``security
proofs''
won't
find
Maxwell's
equations
used
anywhere.
This
justifies
omitting
Maxwell's
equations
as
a
hypothesis,
but
it
also
strongly
suggests
that
the
secret
physical
actions
taken
by
Alice
and
Bob
don't
involve
any
electricity
or
magnetism.
How
could
one
possibly
prove
anything
about
the
effects
of
electromagnetic
actions
without
invoking
the
relevant
laws
of
physics?
Similarly,
Newton's
law
isn't
used
anywhere,
strongly
suggesting
that
the
secret
physical
actions
taken
by
Alice
and
Bob
don't
involve
moving
any
mass
around.
But
then
what
exactly
{\it
are\/}
the
secret
actions
taken
by
Alice
and
Bob?

\section{Abstractions without examples}
A
closer
look
shows
that
the
papers
don't
actually
say
what
physical
actions
Alice
and
Bob
are
hypothesized
to
be
carrying
out.
Sometimes
the
papers
do
specify
physical
details
of,
e.g.,
polarized
photons
being
sent
from
Alice
to
Bob
via
Eve
in
``BB84
QKD'';
or
entangled
pairs
of
photons
being
sent
from
Eve
to
both
Alice
and
Bob
in
``device-independent
E91
QKD''
(see,
e.g.,
\cite{2009/pironio});
or
photons
being
sent
with
particular
timing
in
``relativistic
QKD''
(see,
e.g.,
\cite{2001/molotkov});
but
there
are
always
many
other
steps
taken
by
Alice
and
Bob
whose
physical
details
are
not
mentioned
anywhere
in
any
of
the
``security
proofs''.

Instead
the
papers
make
various
abstract
hypotheses
regarding
quantum
states.
Consider,
for
example,
the
following
hypothesis
from
\cite{2014/masanes}:
\begin{quote}
Eve
can
choose
an
observable
$Z$
and
obtains
an
outcome
$E$.
We
assume
that
this,
together
with
the
public
messages
exchanged
by
Alice
and
Bob,
is
all
information
available
to
her.
\end{quote}
Apparently
Alice
and
Bob
are
supposed
to
take
physical
actions
that
do
not
interact
with
Eve's
quantum
state
except
through
the
public
interactions
specified
by
the
protocol.
This
begs
the
question
of
what
those
physical
actions
are
supposed
to
be.

Obviously
there
are
physical
actions
that
Alice
and
Bob
can
take
in
the
real
world
that
might
look
like
QKD
but
that
don't
actually
meet
the
hypotheses
of
the
QKD
``security
proofs''.
In
particular,
every
QKD
``security
proof''
assumes
that
there
are
various
secret
actions
by
Alice
and
Bob,
actions
unobserved
by
Eve,
such
as
a
sizeable
fraction
of
Alice's
choices
of
polarization
bases
in
the
BB84
protocol.
If
Alice
and
Bob
actually
leak
these
secrets
to
Eve
then
these
hypotheses
are
false
and
the
QKD
``security
proofs''
say
nothing.

Anyone
who
claims
that
the
security
of
QKD
is
guaranteed
by
the
laws
of
physics
is
logically
forced
to
argue
that
these
physical
actions
by
Alice
and
Bob
are
not
actually
QKD:
the
only
physical
actions
that
qualify
as
QKD
are
those
for
which
the
hypotheses
of
the
theorems
are
satisfied.
On
the
other
hand,
many
authors
insist
on
using
the
label
``QKD''
for
actions
that
are
obviously
outside
the
scope
of
the
theorems.
Sometimes
``security''
systems
that
have
already
been
convincingly
demonstrated
to
fail
in
the
real
world
(see,
e.g.,
\cite{2016/huang})
are
still
sold
as
``QKD''.

To
avoid
confusion,
let's
use
the
name
``theoretical
QKD''
to
refer
to
QKD
meeting
all
of
the
hypotheses
of
the
theorems.
Is
it
clear
that
theoretical
QKD
exists
within
the
laws
of
physics?
Are
there
any
examples
of
physical
actions
for
Alice
and
Bob
that
{\it
do\/}
qualify
as
theoretical
QKD?

Consider
the
following
nightmare
scenario
for
the
QKD
``security
proofs'':
The
hypotheses
of
theoretical
QKD
are
actually
{\it
inconsistent
with
the
laws
of
physics}.
For
every
sequence
of
physical
actions
that
Alice
and
Bob
can
take,
the
hypotheses
turn
out
to
be
false.
It
is
vacuous
to
claim
that
the
security
of
theoretical
QKD
is
guaranteed
by
the
laws
of
physics,
since
the
laws
of
physics
imply
that
theoretical
QKD
does
not
exist.
What
evidence
do
we
have
that
the
QKD
``security
proofs''
are
not
in
this
nightmare
scenario?

Part
of
the
job
of
a
mathematician
or
computer
scientist
stating
a
theorem
is
to
verify---or
at
least
plausibly
conjecture---that
there
are
examples
meeting
the
hypotheses
of
the
theorem.
Vacuous
lemmas
are
occasionally
stated
as
intermediate
steps
inside
proofs
by
contradiction,
but
these
lemmas
are
also
clearly
identified
as
being
vacuous,
not
as
saying
anything
meaningful.

\section{The holographic principle}\label{holographic}

Let
me
focus
specifically
on
the
hypothesis
that
Alice
and
Bob
are
taking
various
actions
unobserved
by
Eve.
This
is,
as
I
mentioned
above,
assumed
in
every
QKD
``security
proof''.
Obviously
Alice
and
Bob
have
no
security
against
Eve
if
Eve
observes
all
of
their
secrets,
so
one
cannot
avoid
making
such
a
hypothesis.

This
hypothesis
seems
to
be
flatly
contradicted
by
the
``holographic
principle''.
In
the
words
of
Brian
Greene
\cite[page
272]{2011/greene}:
\begin{quote}
The
{\it
holographic
principle\/}
envisions
that
all
we
experience
may
be
fully
and
equivalently
described
as
the
comings
and
goings
that
take
place
at
a
thin
and
remote
locus.
It
says
that
if
we
could
understand
the
laws
that
govern
physics
on
that
distant
surface,
and
the
way
phenomena
there
link
to
experience
here,
we
would
grasp
all
there
is
to
know
about
reality.
\end{quote}

Readers
not
familiar
with
the
idea
that
the
universe
is
connected
in
this
way
should
consider
radios
as
an
example
of
long-distance
interaction.
Radio
communication
typically
relies
on
Alice
and
Bob
generating
a
reasonably
strong
signal
to
make
the
receiver's
job
easier,
but
this
cooperation
is
not
necessary;
by
building
a
very
large
array
of
radio
receivers,
Eve
can
pick
up
a
very
faint
radio
signal
from
ten
thousand
miles
away.

In
light
of
Maxwell's
equations,
how
can
one
justify
the
notion
that
Eve
is
unable
to
observe
secret
electromagnetic
actions
by
Alice
and
Bob?
Alice
and
Bob
can
try
to
use
a
Faraday
cage
to
block
their
signal,
but
a
Faraday
cage
does
not
create
a
truly
isolated
environment
(see,
e.g.,
\cite{2016/bernstein}
and
\cite{2018/guri});
it
merely
applies
some
scrambling
to
the
signals
emitted
from
that
environment.
One
could
hypothesize
that
Eve
{\it
is
not
observing\/}
the
actions
by
Alice
and
Bob---perhaps
Eve
is
underfunded,
or
simply
lazy---but
this
is
obviously
not
``absolute
security,
guaranteed
by
the
fundamental
laws
of
physics''.

It
seems
that
the
only
way
for
Alice
and
Bob
to
avoid
Maxwell's
information
leak
is
to
avoid
any
data
flow
from
secrets
to
electricity
or
magnetism.
But
then
how
are
Alice
and
Bob
supposed
to
control
their
photon
generators,
photon
detectors,
and
other
QKD
equipment?

More
importantly,
the
holographic
principle
says
that
the
difficulty
here
is
not
limited
to
electromagnetism.
{\it
Any\/}
physical
encoding
of
information
will
be
visible
to
a
sufficiently
resourceful
attacker
watching
signals
on
a
remote
screen.

Well-known
examples
of
signal
processing
(sonar
arrays,
radar
arrays,
X-ray
tomography,
magnetic-resonance
imaging,
etc.)~all
seem
to
fit
the
following
general
rule:
Eve's
cost
for
stealing
a
physically
encoded
secret
grows
only
polynomially
with
Eve's
distance
from
the
secret.
Alice
and
Bob
might
try
to
hide
at
some
reasonable
distance
from
attackers,
and
from
any
attacker-controlled
equipment,
but
this
only
moderately
increases
the
cost
of
the
attack.

QKD
is
claimed
to
have
information-theoretic
security
guaranteed
by
the
laws
of
physics,
not
depending
on
``the
conjectured
difficulty
of
computing
certain
functions''.
Even
if
holographic
signal
processing
actually
has
superpolynomial
cost,
the
availability
of
the
holographic
signal
is
enough
to
contradict
the
QKD
security
claims.
Eve
can
carry
out
all
necessary
computations
at
her
leisure,
retroactively
breaking
QKD
without
having
to
interfere
with
the
protocol
execution.
This
directly
contradicts
the
claims
in
\cite{2013/unruh}
and
\cite[page
17]{2015/schaffner}
that
QKD
offers
``everlasting
security'',
and
the
similar
claim
in
\cite[page
19]{2015/campagna}.

Of
course,
claimed
``laws''
of
physics
do
not
have
the
same
level
of
certainty
as
mathematics.
Many
past
``laws''
seem
inconsistent
with
experiment
and
are
no
longer
believed
to
be
accurate:
e.g.,
Newton's
law
of
gravitation
is
merely
a
first
approximation,
failing
to
take
relativistic
effects
into
account.
One
might
speculate
that
(1)
the
holographic
principle
is
not
true
in
the
level
of
generality
suggested
in
\cite{1993/hooft},
\cite{1995/susskind},
\cite{2002/bousso},
etc.;
(2)
the
leaks
of
information
to
Eve
are
``merely''
through
electromagnetic
waves,
gravity,
etc.;
and
(3)
there
is
some
way
for
Alice
and
Bob
to
carry
out
QKD
without
triggering
any
of
these
leaks
of
information.
In
other
words,
it
is
conceivable
that
some
law
of
physics
not
known
today
will
eventually
guarantee
the
security
of
some
form
of
QKD
not
specified
today.
But
these
unsupported
speculations
are
very
far
from
justifying
the
claim
that
QKD
provides
``absolute
security,
guaranteed
by
the
fundamental
laws
of
physics''.

\section{A concrete attack against QKD}

Here
is
an
attack
to
illustrate
the
principles
of
Section~\ref{holographic}.
An
attacker
turns
on
a
large
array
of
radio
receivers,
leaves
it
on,
and
records
the
resulting
data.
A
tiny
budget,
on
the
scale
of
a
million
dollars
per
year,
will
allow
the
attacker
to
store
a
gigabyte
of
data\footnote
{This
metric
presumes
that
the
radio
signals
are
converted
from
analog
to
digital
before
being
stored.
Perhaps
the
attack
is
even
more
effective
if
the
attacker
records
the
signals
directly
in
some
analog
form.}
per
second.

The
victim
then
begins
using
a
state-of-the-art
QKD
device.
This
device
relies
on
the
secrecy
of
non-quantum
information
encoded
inside
the
equipment
as
conventional
electromagnetic
signals
(not
to
be
confused
with
the
unclonable
quantum
information
sent
via
photons).
These
conventional
electromagnetic
signals
influence
the
data
recorded
by
the
attacker,
as
per
Maxwell's
equations.
This
influence
also
depends
on
other
inputs,
such
as
the
physical
location
and
shape
of
objects
that
have
been
reflecting
radio
waves,
so
the
attacker
also
collects
as
much
information
as
possible
about
these
inputs
by
recording
data
from
a
similarly
large
array
of
cameras,
microphones,
and
other
sensors.

The
attacker
now
computes
the
most
likely
possibilities
for
the
victim's
QKD
secrets,
and
thus
for
the
resulting
secret
keys,
given
the
recorded
data.
This
is
a
mathematical
computation
exploiting
the
laws
of
physics---the
equations
that
determine
the
influence
of
the
QKD
secrets
and
other
inputs
upon
the
data
recorded
by
the
attacker.

From
an
information-theoretic
perspective,
it
seems
likely
that
the
recorded
data
is
sufficient
to
successfully
pinpoint
the
secret
keys
of
all
QKD
users.
There
is
certainly
no
guarantee
to
the
contrary.
There
is
an
algorithmic
challenge
of
optimizing
the
computation
of
these
secrets,
and
perhaps
the
best
algorithm
that
we
can
figure
out
will
take
a
million
years
to
run
on
future
computer
equipment,
but
there
is
again
no
guarantee
of
this.

Now
compare
this
attack
to
the
claim
made
by
Campagna,
Chen,
Dagdelen,
Ding,
Fernick,
Gisin,
Hayford,
Jennewein,
Lütkenhaus,
Mosca,
Neill,
Pecen,
Perlner,
Ribordy,
Schanck,
Stebila,
Walenta,
Whyte,
and
Zhang
in
\cite[page
19]{2015/campagna}:
\begin{quote}
An
important
characteristic
of
quantum
key
distribution
is
that
any
attack
(e.g.~any
attempt
to
exploit
a
flaw
in
an
implementation
of
transmitters
or
receivers)
must
be
carried
out
in
real
time.
Contrary
to
classical
cryptographic
schemes,
in
QKD
there
is
no
way
to
save
the
information
for
later
decryption
by
more
powerful
technologies.
This
greatly
reduces
the
window
of
opportunity
for
performing
an
attack
against
QKD;
the
window
is
much
wider
for
conventional
cryptography.
\end{quote}
This
claim
is
disproven
by
the
QKD
attack
explained
above:
\begin{itemize}
\item
The
attack
does
not
need
to
be
completed
in
``real
time''.
\item
The
attack
does
save
information
for
later
decryption.
\item
The
attack
does
take
advantage
of
more
powerful
technologies
available
in
the
future.
\end{itemize}
The
only
accurate
part
of
the
claim
is
that
the
attack
must
{\it
begin\/}
before
the
victim
uses
QKD---the
attacker
must
plan
ahead
by
collecting
and
storing
data
for
later
analysis---but
the
comparison
in
\cite{2015/campagna}
already
assumes
that
the
attacker
is
collecting
and
storing
data
for
later
analysis.\footnote
{As
a
side
note
on
costs,
it
is
not
obvious
in
advance
exactly
how
much
data
needs
to
be
recorded
for
this
QKD
attack
to
succeed.
This
uncertainty
means
that
a
sensible
attacker
will
opt
for
an
``overkill''
approach
of
recording
as
much
data
as
he
can
afford
to
collect;
subsequent
analysis
will
show
whether
this
was
enough,
and
the
attacker
will
then
use
this
experience
to
optimize
subsequent
attacks.
For
comparison,
an
attacker
who
records
the
conventional
encoding
of
all
data
sent
through
an
Internet
router
is
confidently
recording
all
data
that
normal
users
of
the
router
will
act
upon.}
In
the
real
world,
ever-increasing
amounts
of
data
are
being
stored
from
a
wide
range
of
collection
devices;
in
other
words,
the
attack
has
already
begun.

Someone
who
claims
that
the
security
of
QKD
is
``guaranteed
by
the
laws
of
physics''
is
logically
forced
to
claim
that
the
laws
of
physics
guarantee
that
this
particular
attack
does
not
work.
When
I
ask
such
people
why
they
think
that
this
attack
will
fail,
they
typically
answer
in
one
of
the
following
ways:
\begin{itemize}
\item
Speculation:
Claim
that
the
electromagnetic
signal
is
too
weak
for
the
attacker
to
pick
up
at
distance
$x$.

This
might
be
true
in
some
cases,
depending
on
$x$
and
on
many
more
details
of
what
the
attacker
is
doing,
but
all
of
this
needs
analysis---see
Section~\ref{countermeasures}.
More
to
the
point,
even
in
situations
where
this
claim
is
true,
this
is
not
the
same
as
security
``guaranteed
by
the
laws
of
physics''.
\item
Abstraction:
Describe
today's
electromagnetic
devices
as
mere
``implementations''
of
an
abstract
concept
of
theoretical
QKD;
claim
that
any
exploitable
``imperfections''
in
these
``implementations''
will
be
corrected
by
future
``implementations''.

This
claim
begs
the
question
of
what
exactly
those
``implementations''
will
be:
how
can
theoretical
QKD
be
embedded
into
the
physical
world,
in
light
of
the
holographic
principle?
More
to
the
point,
this
claim
implicitly
admits
that
the
QKD
security
``guarantee''
is
not
actually
from
the
laws
of
physics,
but
rather
from
a
simplified---I
say
oversimplified---model.
\item
Marketing:
Outline
particular
examples
of
next-generation
``device-independent''
E91
QKD
devices;
say
that
it
is
important
to
fund
development
of
these
devices;
and
claim
that
{\it
these\/}
QKD
devices
will
have
security
guaranteed
by
the
laws
of
physics
where
previous
BB84
QKD
devices
did
not.

At
this
point
the
claims
are
not
even
marginally
consistent
with
the
literature,
which
claims
security
guarantees
for
{\it
both\/}
BB84
and
E91.
\item
Distraction:\footnote{Hey,
look,
a
squirrel!}
Complain
that
other
forms
of
cryptography
are
also
potentially
vulnerable
to
this
type
of
attack.

This
complaint
is
both
correct
and
irrelevant.
What
I
am
disputing
is
the
frequent
unjustified
claim
$Q$
that
quantum
cryptography
has
security
guaranteed
by
the
laws
of
physics.
Perhaps
somewhere
there
is
some
other
unjustified
claim
$C$
that
some
other
form
of
cryptography
has
security
guaranteed
by
the
laws
of
physics,
but
it
is
illogical
to
use
a
complaint
about
$C$
as
a
justification
for
$Q$.
\end{itemize}
I
do
not
have
experimental
evidence
that
this
particular
QKD
attack
works,
but
I
also
see
no
justification
for
the
claim
that
the
attack
will
fail.

\section{Previous work}

There
is
a
huge
literature
claiming
that
the
security
of
theoretical
QKD
is
guaranteed
by
the
laws
of
physics.
I
am
not
aware
of
previous
papers
clearly
and
directly
raising
the
possibility
that
this
claim
is
vacuous---that
theoretical
QKD
is
not
actually
compatible
with
the
laws
of
physics.

Plaga
in
\cite{2006/plaga}
pointed
out
that
speculations
about
nonlinearity
of
quantum
gravity,
if
correct,
could
allow
Eve
to
extract
information
from
the
data
{\it
explicitly
communicated\/}
in
some
QKD
protocols.
What
I
am
saying
is
quite
different:
I
am
focusing
on
information
leaked
from
the
``secret''
portions
of
QKD
protocols,
and
I
am
relying
on
core
physical
phenomena
such
as
electromagnetism.

More
relevant
is
a
paper
\cite{2002/rudolph}
by
Rudolph,
which
(among
other
things)
questions
the
``impenetrability/no
emanation''
hypotheses
for
theoretical
QKD.
What
I
am
saying
is
stronger
in
three
ways:
\begin{itemize}
\item
Rudolph
merely
questions
the
hypotheses,
while
I
am
questioning
both
the
hypotheses
and
the
conclusion.
\item
Regarding
details,
Rudolph
touches
upon
ways
that
{\it
some\/}
secrets
could
leak,
and
asks
whether
these
leaks
can
be
limited
to
``an
exponentially
small
amount
of
useful
information'',
while
I
am
pointing
to
ways
that
{\it
all\/}
secrets
leak
to
a
sufficiently
resourceful
attacker.
\item
Rudolph
says
that
a
``black
hole
lab''
would
``presumably''
satisfy
the
hypotheses,
while
I
am
not
drawing
any
such
line.
My
impression
of
the
consensus
of
physicists
is
that
black
holes
are
{\it
not\/}
an
exception
to
the
holographic
principle.
This
is
important
because
the
extreme
case
of
a
``black
hole
lab''
provides
some
intuition
for
the
best
that
one
can
hope
for
by
adding
more
and
more
shielding.
\end{itemize}
Rudolph
is
saying
that
secrets
are
not
{\it
exactly\/}
unobservable,
while
I
am
giving
reasons
to
believe
that
secrets
are
not
even
{\it
approximately\/}
unobservable,
and
that
the
answer
to
Rudolph's
``exponentially
small''
question
is
negative.
This
is
important
for
claims
that
QKD
is
improving,
eliminating
``imperfections'',
etc.

Brassard
stated
in
\cite{2006/brassard}
that
the
original
prototype
implementation
of
QKD
``was
unconditionally
secure
{\it
against
any
eavesdropper
who
happened
to
be
deaf\/}!
{\tt
:-)}''
(italics
and
smiley
in
original).
This
is
still
an
overstatement
of
the
security
provided
by
the
implementation.
It
is
clear
that
the
implementation
would
also
have
given
up
all
of
its
secrets
to,
e.g.,
a
deaf
eavesdropper
watching
the
screen
of
an
oscilloscope
attached
to
a
small
coil
of
wire.
More
importantly,
Brassard
presented
this
security
failure
as
being
specific
to
{\it
one\/}
implementation
of
QKD,
not
recognizing
the
possibility
of
the
laws
of
physics
forcing
the
same
fundamental
type
of
security
failure
to
appear
in
{\it
all\/}
forms
of
QKD.

A
series
of
papers
on
``quantum
hacking'',
such
as
\cite{2016/huang}
by
Huang,
Sajeed,
Chaiwongkhot,
Soucarros,
Legre,
and
Makarov,
have
broken
the
security
of
various
commercial
implementations
of
QKD\null.
All
of
these
attacks
follow
the
theme
of
Eve
interacting
physically
(passively
or
actively)
with
the
``secret''
computations
by
Alice
and
Bob.

Various
earlier
attacks
are
surveyed
by
Scarani
and
Kurtsiefer
in
\cite{2014/scarani},
who
generally
express
skepticism
regarding
the
security
of
QKD
but
also
claim
that
``in
principle
QKD
can
be
made
secure''.
A
careful
reading
shows
that
this
claim
again
starts
from---and
does
not
attempt
to
justify---the
questionable
hypothesis
that
Alice
and
Bob
are
taking
various
actions
unobserved
by
Eve.
This
hypothesis
is
obviously
false
in
various
{\it
examples\/}
of
successful
QKD
attacks
summarized
in
\cite{2014/scarani};
but
Scarani
and
Kurtsiefer,
like
Brassard,
do
not
recognize
the
possibility
of
the
holographic
principle
allowing
Eve
to
violate
the
hypothesis
for
{\it
all\/}
possible
QKD
devices.

None
of
the
previous
literature
has
stopped
researchers
from
claiming
that
the
security
of
QKD
is
guaranteed
by
the
laws
of
physics---and
dismissing
every
QKD
security
failure,
every
leak
of
secrets,
as
a
supposedly
fixable
implementation
mistake.

\section{Countermeasures}\label{countermeasures}

What
should
Alice
and
Bob
do
if
the
promise
of
``security
guaranteed
by
the
fundamental
laws
of
physics''
is
a
sham---in
particular,
if
physical
effects
allow
a
computationally
unlimited
attacker
to
steal
secrets
from
an
arbitrary
distance?

The
obvious
answer
is
to
study
the
{\it
cost\/}
of
Eve's
attack,
and
to
take
measures
to
increase
this
cost---hopefully
beyond
anything
that
Eve
can
afford.
There
are
several
ways
to
argue
that
Eve
is
subject
to
cost
limits,
such
as
the
following:
\begin{itemize}
\item
Perhaps
the
lifetime
of
the
universe
is
limited.
\item
Perhaps,
even
in
an
everlasting
universe,
the
cosmological
constant
is
positive,
putting
a
limit
on
all
computations,
as
explained
in
\cite{2000/bousso}.
This
constant
is
generally
believed
to
be
around
$10^{-122}$.
\item
Perhaps
the
space
aliens
who
control
most
of
the
resources
in
the
universe
are
happy
that
Alice
and
Bob
are
secretly
arranging
climate-change
protests,
and
are
not
willing
to
help
policewoman
Eve
see
these
secrets.
Compare
\cite{2015/neslen}.
\end{itemize}
If
Eve
is
limited
to
cost
$C$,
then
Alice
and
Bob
do
not
need
to
aim
for
the
unachievable
goal
of
security
guaranteed
by
the
laws
of
physics;
Alice
and
Bob
merely
need
to
be
secure
against
all
attacks
that
cost
at
most
$C$.

Earlier
I
mentioned
that
Eve's
cost
for
stealing
a
physically
encoded
secret
seems
to
grow
only
polynomially
with
Eve's
distance
from
the
secret.
However,
this
polynomial
also
seems
to
depend
heavily
on
the
choice
of
encoding
mechanism.
Alice
and
Bob
can
and
should
choose
technologies
for
encoding
their
secrets
with
the
goal
of
making
this
polynomial
as
large
as
possible.
For
example,
one
might
guess
that
the
following
steps
help:
\begin{itemize}
\item
Store
and
process
secrets
inside
a
modern
10nm
Intel
CPU,
rather
than
using
older,
less
efficient
chip
technology.
Presumably
consuming
less
energy
for
the
same
computation
will
reduce
the
amount
of
signal
visible
to
Eve,
increasing
Eve's
cost
for
recovering
the
secrets.
\item
Hide
the
secrets
by
adding
shields
and
by
generating
extra
noise.
A
Faraday
cage
leaks
information,
as
mentioned
above,
but
seems
to
make
the
information
more
difficult
to
intercept
and
decipher.
\item
Mask
the
secrets,
for
example
by
encoding
bit
0
as
a
random
even-weight
4-bit
string
and
encoding
bit
1
as
a
random
odd-weight
4-bit
string.
\item
Apply
more
sophisticated
mathematical
encodings,
such
as
``secret-key
cryptography''
and
``public-key
cryptography''.
\end{itemize}
Of
course,
guessing
is
not
the
same
as
systematically
studying
the
actual
impact
on
attack
cost.
The
premier
venue
for
scientific
papers
analyzing
the
costs
of
attackers
learning
secrets
from
(passive
and
active)
physical
effects,
and
analyzing
the
costs
of
users
defending
themselves
against
these
attacks,
is
the
``Cryptographic
Hardware
and
Embedded
Systems''
(CHES)
conference
series.
CHES
has
run
every
year
since
1999
and
has
attracted
more
than
300
attendees
every
year
since
2009.

I
can
easily
be
accused
of
bias---I've
served
on
the
CHES
program
committee
every
year
since
2008
(looking
mainly
at
the
costs
for
users)---but
I
don't
think
anyone
can
dispute
the
need
for
this
type
of
research.
Competent
attackers
take
advantage
of
not
merely
the
information
that
we
declare
as
public
but
also
all
of
the
information
they
can
see
through
every
available
side
channel.
We
have
to
take
the
same
perspective,
taking
account
of
all
aspects
of
how
secrets
are
embedded
into
the
real
world,
if
we
want
to
build
information-protection
systems
that
society
can
afford
to
use
but
that
the
attackers
find
infeasible
to
break.

The
QKD
literature
doesn't
try
to
argue
that
QKD
will
produce
improvements
within
the
traditional
chart
of
$(x,y)=(\hbox{user
cost},\hbox{attack
cost})$.
Instead
the
QKD
literature
tries
to
dodge
cost
questions
by
claiming
that
QKD
inhabits
a
magical
realm
beyond
the
top
of
the
chart---security
``guaranteed
by
the
laws
of
physics''.
Unfortunately,
this
claim
is
not
justified
anywhere
in
the
literature,
and
it
seems
very
difficult
to
justify,
in
light
of
what
the
laws
of
physics
actually
say.

\begingroup
\catcode`@=11
\def\cite#1{\relax
\mypdfdest{citing-\the\backciteanchor}\relax
\@tempswafalse\protect\@mycitex[]{#1}\relax
\immediate\write\@auxout{\string\mybackbibcite{#1}{[\pdfstartlink
attr{/Border[0
0
0]/H/I}
goto
name{citing-\the\backciteanchor}\string\textcolor{red}{\the\referencenumber}\string\relax\pdfendlink]}}\relax
\expandafter\protected@xdef\csname mybackbibcite@\refno\endcsname{\relax}\relax
\global\advance\backciteanchor1\relax
}

\def\bib#1{\catcode`\^^M=13\def\refno{#1}
\let\refeditors\relax
\let\refbyone\relax\let\refbytwo\relax\let\refbythree\relax
\let\refpaper\relax\let\refjour\relax\let\refpaperyr\relax
\let\refvol\relax\let\refpages\relax
\let\refpaperinfo\relax
\let\refinbook\relax
\let\refbook\relax\let\refbookyr\relax
\let\refedition\relax\let\refprinting\relax
\let\refthesis\relax
\let\refseries\relax
\let\refseriesvol\relax
\let\refbookinfo\relax
\let\refpubl\relax\let\refpubladdr\relax
\let\refyr\relax
\let\refurl\relax\let\refid\relax
\let\refdoi\relax
\let\refolder\relax
\let\refnewer\relax
\let\refalso\relax
\let\refmr\relax\let\refisbn\relax\let\refissn\relax}
\catcode`\^^M=13\def^^M{}
\def\by#1^^M{\ifx\refbyone\relax
\def\refbyone{#1}\else\ifx\refbytwo\relax
\def\refbytwo{#1}\else\ifx\refbythree\relax
\def\refbythree{#1}\else\let\refbyold\refbyone
\edef\refbyone{\refbyold, \refbytwo}\let\refbytwo\refbythree
\def\refbythree{#1}\fi\fi\fi}
\def\editor#1^^M{\ifx\refbyone\relax
\def\refbyone{#1}\def\refeditors{ (editor)}\else\ifx\refbytwo\relax
\def\refbytwo{#1}\def\refeditors{ (editors)}\else\ifx\refbythree\relax
\def\refbythree{#1}\else\let\refbyold\refbyone
\edef\refbyone{\refbyold, \refbytwo}\let\refbytwo\refbythree
\def\refbythree{#1}\fi\fi\fi}
\def\book#1^^M{\def\refbook{, \it #1\rm}}
\def\paper#1^^M{\def\refpaper{, \it #1\rm}}
\def\yr#1^^M{\def\refpaperyr{ (#1)}\def\refbookyr{, #1}}
\def\edition#1^^M{\def\refedition{, #1 edition}}
\def\citeas#1^^M{}
\def\citeyr#1^^M{}
\def\printing#1^^M{\def\refprinting{, #1 printing}}
\def\phdthesis^^M{\def\refthesis{, Ph.D. thesis}}
\def\mathesis^^M{\def\refthesis{, M.A. thesis}}
\def\series#1^^M{\def\refseries{, #1}}
\def\seriesvol#1^^M{\def\refseriesvol{, #1}}
\def\bookinfo#1^^M{\def\refbookinfo{, #1}}
\def\publ#1^^M{\def\refpubl{, #1}}
\def\publaddr#1^^M{\def\refpubladdr{, #1}}
\def\seeolder#1^^M{\def\refolder{; see also older version #1}}
\def\seenewer#1^^M{\def\refnewer{; see also newer version #1}}
\def\also#1^^M{\def\refalso{; #1}}
\def\inbook#1^^M{\def\refinbook{, in #1}}
\def\paperinfo#1^^M{\def\refpaperinfo{, #1}}
\def\jour#1^^M{\def\refjour{, #1}}
\def\vol#1^^M{\def\refvol{ \bf#1\rm}}
\def\mr#1^^M{\def\refmr{. MR #1}} \def\isbn#1^^M{\def\refisbn{. ISBN #1}}
\def\issn#1^^M{\def\refissn{. ISSN #1}}
\def\url#1^^M{\def\refurl{. URL: \showurl{#1}}}
\def\doi#1^^M{\def\refdoi{. DOI: \doilink{#1}}}
\def\id#1^^M{\def\refid{. ID \def\next{}\showid #1^^A^^A}}
\def\pages#1^^M{\def\refpages{, #1}}
\catcode`\^^M=5

\def\showid{\ifx\next^^A\else{\tt\next\penalty500}\afterassignment\showid\fi\let\next=}

\newcount\referencenumber
\def\thereferencenumber{[\the\referencenumber]}

\def\endref{\catcode`\^^M=5{}\relax
\advance\referencenumber1\relax
\immediate\write\@auxout{\string\bibcite{\refno}{\the\referencenumber}}
\item[\thereferencenumber]
{\noindent\hbox
to
0pt{\hss\mypdfdest{bib-\refno}\hskip
20pt}\relax
\ifx\refbyone\relax--- (no editor)\else\refbyone\ifx\refbytwo\relax\else
\ifx\refbythree\relax, \refbytwo\else, \refbytwo, \refbythree\fi\fi\fi
\refeditors
\ifx\refpaper\relax\else
\refpaper\refinbook\refpaperinfo
\refjour\refvol\refpaperyr\refpages
\fi\relax
\ifx\refbook\relax\else
\refbook\refedition\refprinting
\refthesis\refseries\refseriesvol\refbookinfo
\refpubl\refpubladdr\refbookyr
\fi\relax
\refolder\refnewer\refalso
\refisbn\refissn\refmr\refurl\refdoi\refid
\expandafter\ifx\csname mybackcite@\refno\endcsname\relax
\else\relax
\csname mybackcite@\refno\endcsname
\fi\relax
\expandafter\ifx\csname mybackbibcite@\refno\endcsname\relax
\else\relax
\csname mybackbibcite@\refno\endcsname
\fi\relax
.}\ignorespaces}

\catcode`#=11
\catcode`@=12
\catcode`~=11

\endgroup


\begin{thebibliography}{[99]}

\bib{2006/-awaji}
\yr 2006
\book Proceedings of IEEE information theory workshop on theory and practice in information theoretic security, Awaji Island, Japan, October 2005
\endref

\bib{1993/salamfestschrift}
\yr 1993
\editor Ahmed Ali
\editor John Ellis
\editor Seifallah Randjbar-Daemi
\book Salamfestschrift: a collection of talks from the Conference on Highlights of Particle and Condensed Matter Physics, ICTP, Trieste, Italy, 8--12 March 1993
\series World Scientific Series in 20th Century Physics
\seriesvol 4
\publ World Scientific
\endref

\bib{2016/bernstein}
\yr 2016
\by Daniel J. Bernstein
\paper The inverse Faraday challenge (talk slides)
\url https://cr.yp.to/talks.html#2016.08.18
\endref

\bib{2000/bousso}
\yr 2000
\by Raphael Bousso
\paper Positive vacuum energy and the $N$-bound
\jour Journal of High Energy Physics
\vol 11
\pages 038
\url https://arxiv.org/abs/hep-th/0010252
\doi 10.1088/1126-6708/2000/11/038
\endref

\bib{2002/bousso}
\yr 2002
\by Raphael Bousso
\paper The holographic principle
\url https://arxiv.org/abs/hep-th/0203101
\jour Reviews of Modern Physics
\vol 74
\pages 825--874
\doi 10.1103/RevModPhys.74.825
\endref

\bib{2006/brassard}
\yr 2006
\by Gilles Brassard
\paper Brief history of quantum cryptography: a personal perspective
\pages 19--23
\inbook \cite{2006/-awaji}
\url https://arxiv.org/abs/quant-ph/0604072
\doi 10.1109/ITWTPI.2005.1543949
\endref

\bib{2015/campagna}
\yr 2015
\by Matthew Campagna
\by Lidong Chen
\by \"Ozgür Dagdelen
\by Jintai Ding
\by Jennifer K. Fernick
\by Nicolas Gisin
\by Donald Hayford
\by Thomas Jennewein
\by Norbert Lütkenhaus
\by Michele Mosca
\by Brian Neill
\by Mark Pecen
\by Ray Perlner
\by Grégoire Ribordy
\by John M. Schanck
\by Douglas Stebila
\by Nino Walenta
\by William Whyte
\by Zhenfei Zhang
\paper Quantum safe cryptography and security: an introduction, benefits, enablers and challenges
\paperinfo ETSI White Paper No. 8
\url http://www.etsi.org/images/files/ETSIWhitePapers/QuantumSafeWhitepaper.pdf
\endref

\bib{2013/canetti}
\yr 2013
\editor Ran Canetti
\editor Juan A. Garay
\book Advances in cryptology---CRYPTO 2013---33rd annual cryptology conference, Santa Barbara, CA, USA, August 18--22, 2013, proceedings, part I
\series Lecture Notes in Computer Science
\seriesvol 8042
\publ Springer
\endref

\bib{2004/christandl}
\yr 2004
\by Matthias Christandl
\by Renato Renner
\by Artur Ekert
\paper A generic security proof for quantum key distribution
\url https://arxiv.org/abs/quant-ph/0402131
\endref

\bib{2004/gottesman}
\yr 2004
\by Daniel Gottesman
\by Hoi-Kwong Lo
\by Norbert L\"utkenhaus
\by John Preskill
\paper Security of quantum key distribution with imperfect devices
\jour Quantum Information and Computation
\vol 4
\pages 325--360
\url https://arxiv.org/abs/quant-ph/0212066
\endref

\bib{2011/greene}
\yr 2011
\by Brian Greene
\book The hidden reality: Parallel universes and the deep laws of the cosmos
\publ Penguin
\isbn 9781846145353
\endref

\bib{2018/guri}
\yr 2018
\by Mordechai Guri
\by Andrey Daidakulov
\by Yuval Elovici
\paper MAGNETO: Covert channel between air-gapped systems and nearby smartphones via CPU-generated magnetic fields
\url https://arxiv.org/abs/1802.02317
\endref

\bib{1993/hooft}
\yr 1993
\by Gerard 't Hooft
\paper Dimensional reduction in quantum gravity
\inbook \cite{1993/salamfestschrift}
\url https://arxiv.org/abs/gr-qc/9310026
\endref

\bib{2016/huang}
\yr 2016
\by Anqi Huang
\by Shihan Sajeed
\by Poompong Chaiwongkhot
\by Mathilde Soucarros
\by Matthieu Legre
\by Vadim Makarov
\paper Gaps between industrial and academic solutions to implementation loopholes in QKD: testing random-detector-efficiency countermeasure in a commercial system
\url https://arxiv.org/abs/1601.00993
\doi 10.1109/JQE.2016.2611443
\endref

\bib{1995/hughes}
\yr 1995
\by Richard J. Hughes
\by Douglas M. Alde
\by P. Dyer
\by Gabriel G. Luther
\by George L. Morgan
\by Martin M. Schauer
\paper Quantum cryptography
\jour Contemporary Physics
\vol 36
\url https://arxiv.org/abs/quant-ph/9504002
\doi 10.1080/00107519508222149
\pages 149--163
\endref

\bib{2016/idq-understanding}
\yr 2016
\by ID Quantique
\paper Understanding quantum cryptography, version 3.0
\url https://tinyurl.com/y9zok2c2
\endref

\bib{2014/masanes}
\yr 2014
\by Llu\'is Masanes
\by Renato Renner
\by Matthias Christandl
\by Andreas Winter
\by Jonathan Barrett
\paper Full security of quantum key distribution from no-signaling constraints
\jour IEEE Transactions on Information Theory
\vol 60
\pages 4973--4986
\url https://arxiv.org/abs/quant-ph/0606049
\doi 10.1109/TIT.2014.2329417
\endref

\bib{2001/molotkov}
\yr 2001
\by Sergey N. Molotkov
\by Sergey S. Nazin
\paper A simple proof of the unconditional security of relativistic quantum cryptography
\jour Journal of Experimental and Theoretical Physics
\vol 92
\pages 871--878
\url https://arxiv.org/abs/quant-ph/0008008
\doi 10.1134/1.1378181
\endref

\bib{2015/neslen}
\yr 2015
\by Arthur Neslen
\paper Paris climate activists put under house arrest using emergency laws
\url http://www.theguardian.com/environment/2015/nov/27/paris-climate-activists-put-under-house-arrest-using-emergency-laws
\endref

\bib{2009/pironio}
\yr 2009
\by Stefano Pironio
\by Antonio Ac\'in
\by Nicolas Brunner
\by Nicolas Gisin
\by Serge Massar
\by Valerio Scarani
\paper Device-independent quantum key distribution secure against collective attacks
\jour New Journal of Physics
\vol 11
\pages 045021 (25pp)
\url http://www.unige.ch/gap/quantum/_media/publications:bib:qkddeviceindepfull.pdf
\doi 10.1088/1367-2630/11/4/045021
\endref

\bib{2006/plaga}
\yr 2006
\by Rainer Plaga
\paper A fundamental threat to quantum cryptography: gravitational attacks
\jour The European Physical Journal D---Atomic, Molecular, Optical and Plasma Physics
\vol 38
\pages 409--413
\url http://link.springer.com/article/10.1140/epjd/e2006-00045-y
\endref

\bib{2002/rudolph}
\yr 2002
\by Terry Rudolph
\paper The laws of physics and cryptographic security
\url https://arxiv.org/abs/quant-ph/0202143
\endref

\bib{2014/scarani}
\yr 2014
\by Valerio Scarani
\by Christian Kurtsiefer
\paper The black paper of quantum cryptography: Real implementation problems
\jour Theoretical Computer Science
\vol 560
\pages 27--32
\url https://arxiv.org/abs/0906.4547
\endref

\bib{2015/schaffner}
\yr 2015
\by Christian Schaffner
\paper Quantum cryptography: from key distribution to position-based cryptography (talk slides)
\url https://events.ccc.de/congress/2015/Fahrplan/events/7305.html
\endref

\bib{2000/shor}
\yr 2000
\by Peter W. Shor
\by John Preskill
\paper Simple proof of security of the BB84 quantum key distribution protocol
\jour Physical Review Letters
\vol 85
\pages 441--444
\url https://arxiv.org/abs/quant-ph/0003004
\doi 10.1103/PhysRevLett.85.441
\endref

\bib{1995/susskind}
\yr 1995
\by Leonard Susskind
\paper The world as a hologram
\jour Journal of Mathematical Physics
\vol 36
\pages 6377--6396
\url https://arxiv.org/abs/hep-th/9409089
\doi 10.1063/1.531249
\endref

\bib{2013/unruh}
\yr 2013
\by Dominique Unruh
\paper Everlasting multi-party computation
\pages 380--397
\inbook Crypto 2013 \cite{2013/canetti}
\url https://eprint.iacr.org/2012/177
\endref

\end{thebibliography}
\end{document}